\documentclass{eas}
\usepackage{graphicx}
\newcommand{\Ps}{ P_\star}

\newcommand{\Porb}{ P_{\rm orb}}

\newcommand{\Msun}{ M_\odot}

\newcommand{\Mjup}{ M_{\rm Jup}}

\usepackage[normalem]{ulem}
\usepackage{color}
\definecolor{blue}{RGB}{0,0,255}
\definecolor{red}{RGB}{255,0,0}
\definecolor{green}{RGB}{0,200,0}
\definecolor{black}{RGB}{0,0,0}

\begin{document}

\title{Effect of the rotation, tidal dissipation history and metallicity of stars on the evolution of close-in planets} 
\author{Emeline Bolmont}\address{NaXys, Department of Mathematics, University of Namur, 8 Rempart de la Vierge, 5000 Namur, Belgium}
\author{Florian Gallet}\address{Department of Astronomy, University of Geneva, Chemin des Maillettes 51, 1290 Versoix, Switzerland}
\author{St\'ephane Mathis}\address{Laboratoire AIM Paris-Saclay, CEA/Irfu Universit\'e Paris-Diderot CNRS/INSU, 91191 Gif-sur-Yvette, France}
\author{Corinne Charbonnel$^2$}
\author{Louis Amard$^2$}
\begin{abstract}
Since 1995, numerous close-in planets have been discovered around low-mass stars (M to A-type stars).
These systems are susceptible to be tidally evolving, in particular the dissipation of the kinetic energy of tidal flows in the host star may modify its rotational evolution and also shape the orbital architecture of the surrounding planetary system.
Recent theoretical studies have shown that the amplitude of the stellar dissipation can vary over several orders of magnitude as the star evolves, and that it also depends on the stellar mass and rotation.
We present here one of the first studies of the dynamics of close-in planets orbiting low-mass stars (from $0.6~\Msun$ to $1.2~\Msun$) where we compute the simultaneous evolution of the star's structure, rotation and tidal dissipation in its external convective envelope.
We demonstrate that tidal friction due to the stellar dynamical tide, i.e. tidal inertial waves (their restoring force is the Coriolis acceleration) excited in the convection zone, can be larger by several orders of magnitude than the one of the equilibrium tide currently used in celestial mechanics.
This is particularly true during the Pre Main Sequence (PMS) phase and to a lesser extent during the Sub Giant (SG) phase.
Numerical simulations show that only the high dissipation occurring during the PMS phase has a visible effect on the semi-major axis of close-in planets.
We also investigate the effect of the metallicity of the star on the tidal evolution of planets.
We find that the higher the metallicity of the star, the higher the dissipation and the larger the tidally-induced migration of the planet.

\end{abstract}
\maketitle
\section{Introduction}

The motivation for this study is to make a first step in the combination of advances obtained by two communities working on different aspects of tides. 

On the one hand, a wide part of the community working on the tidal orbital evolution of planetary systems uses an equilibrium tide model (such as the constant time lag model: Mignard~\cite{Mignard1979},  Hut~\cite{Hut1981}, Eggleton~\etal~\cite{EKH1998}). 
This model allows for fast computation, works for all eccentricities and thus facilitates a wide variety of tidal orbital evolution studies (Leconte~\etal~\cite{Leconte2010}, Bolmont~\etal~\cite{Bolmont2011}, \cite{Bolmont2012}, \cite{Bolmont2015}). 
Usually in such a tidal model, the tidal dissipation is taken to be a constant and calibrated on observations (Hansen~\cite{Hansen2010}, \cite{Hansen2012}).
This model is also generally used for rocky planets, however it has been shown that the tidal response of such bodies could be very different thus leading to distinct tidal and rotational evolutions (see Efroimsky \& Lainey~\cite{EfroimskyLainey2007}, Mathis \& Le Poncin-Lafitte~\cite{Mathis2009}, Remus~\etal~\cite{Remus2012a}, Makarov \& Efroimsky \cite{Makarov2013}, Auclair-Desrotour~\etal~\cite{Auclair-Desrotour2014}).

On the other hand, the community specialized in computing the tidal response of bodies uses complex models that are not compatible yet with the need of fast orbital calculations.
However, these models take into account the fact that the stellar dissipation depends on the parameters of the stars and therefore evolves with time. 
For instance, during the lifetime of a star and especially during the PMS phase, its internal structure, radius and rotation evolve drastically.
During the PMS phase and for stars with solar metallicity and with masses between $0.35~\Msun$ and $1.4~\Msun$, a radiative core appears and gets bigger while the size of the convective enveloppe decreases. 
In the meantime, the rotation of the star evolves drastically: it spins up on the PMS phase and then spins down due to the stellar wind (e.g., Gallet \& Bouvier~\cite{GalletBouvier2013}, \cite{GalletBouvier2015}).
All these structural and rotational changes impact the way the star dissipates the kinetic energy of the tidal flows (e.g., Zahn~\cite{Zahn1966}, \cite{Zahn1975}, \cite{Zahn1977}, \cite{Zahn1989}, Ogilvie \& Lin \cite{OgilvieLin2007}, Remus, Mathis and Zahn \cite{Remus2012b}, Mathis~\cite{Mathis2015}, Mathis~\etal~\cite{Mathis2016}).

\section{Model}\label{model}

\begin{figure}[htbp!]
        \begin{center}
        \includegraphics[width=\linewidth]{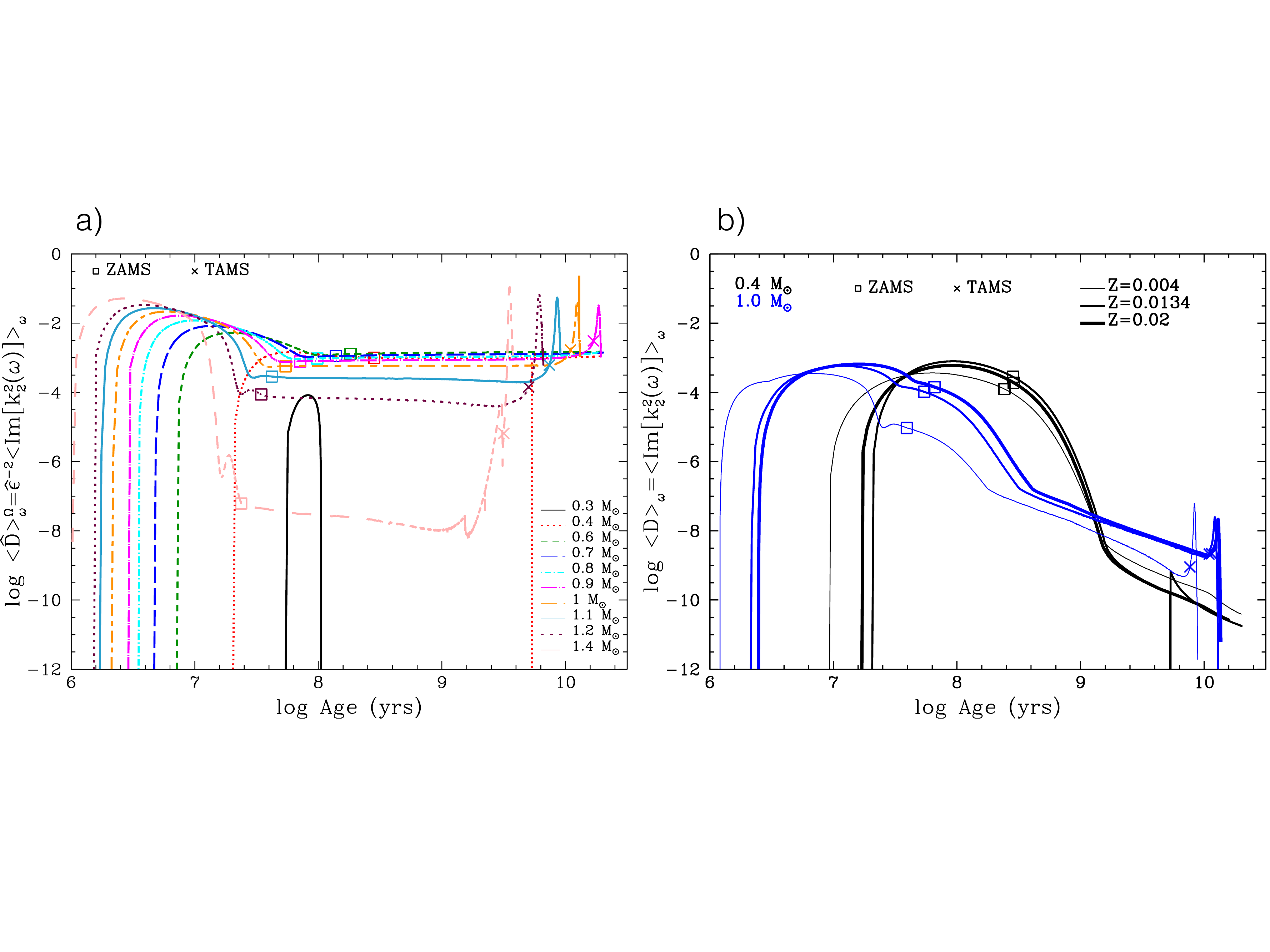}
        \caption{Evolution of the tidal inertial waves dissipation in the external convective envelope of stars with different masses at fixed rotation. a) Tidal dissipation of stars with masses between $0.3~\Msun$ and $1.4~\Msun$, for solar metallicity and constant stellar spin (Figure from Gallet~\etal~in prep). b) Tidal dissipation of stars of mass 0.4 and $1.0~\Msun$ for three different metallicities: Z = 0.004, Z = 0.0134 (Z$_\odot$) and Z = 0.0255, and for an evolving spin.
        The squares represent the Zero Age Main Sequence (ZAMS) and the crosses the Terminal Age Main Sequence (TAMS). Between squares and crosses, the stars are on the MS.}
        \label{dissipM_dissipZ}
        \end{center}
        \end{figure}

We use here the model introduced in Bolmont~\&~Mathis~(\cite{Bolmont2016}), which takes into account the evolution of the tidal dissipation computed by Mathis~(\cite{Mathis2015}) following a first theoretical prescription derived by Ogilvie~(\cite{Ogilvie2013}).
We refer the reader to these articles for a complete discussion of the assumptions.
In this model, a frequency-averaged tidal dissipation in the convective envelope of low-mass stars is computed.
The source of the dissipation is the convective turbulent friction applied on tidal inertial waves. 
These waves are driven by the Coriolis acceleration. 
In presence of a stably stratified core, they can form sheared structures where dissipation is important (Ogilvie \& Lin~\cite{OgilvieLin2007}, Goodman \& Lackner~\cite{GoodmanLackner2009}).
However, the models of Bolmont~\&~Mathis~(\cite{Bolmont2016}) were computed only up to the end of the MS.
We use in this study new grids of tidal dissipation computed with the evolution code STAREVOL (Gallet~\etal,~submitted), which follows the stellar evolution to the SG phase and even the Red Giant Branch (RGB) phase for the higher masses considered ($0.9$ -- $1.4~\Msun$).
Figure \ref{dissipM_dissipZ}a) shows the evolution of the tidal dissipation $\left<{\mathcal D}\right>_{\omega}^\Omega$ when assuming a constant stellar spin $\Omega$ (as done in Eqs. 1 and 4 of Mathis~\cite{Mathis2015}). 
From the PMS phase to the beginning of the MS phase, the tidal dissipation decreases by two orders of magnitude as the stellar radius decreases and the radius and mass aspect ratios of the convective envelope increase (e.g., Gallet~\etal,~submitted).
It then stays relatively constant on the MS as the stellar radius and the aspect ratios vary modestly compared to the PMS and to the post-MS phases.
From the MS turnoff to the SG phase, the stellar radius and the aspect ratios evolve importantly once more and lead first to an increase of the tidal dissipation by two orders of magnitude and second to its decrease towards the RGB phase.

Figure~\ref{dissipM_dissipZ}b) shows the evolution of the tidal dissipation $\left<{\mathcal D}\right>_{\omega}$, which scales as $\Omega^2$, for an evolving stellar spin, for different metallicities and for two stellar masses.
As the star evolves, it first spins up due to the radius contraction and then spin down from the ZAMS due to the magnetic braking induced by stellar winds (here computed according to Bouvier~\etal~\cite{Bouvier1997}. See also Bouvier~\etal~\cite{Bouvier2008}, Gallet \& Bouvier \cite{GalletBouvier2015}, Amard~\etal~\cite{Amard2016}).
From now on, we designate by dissipation the frequency-averaged dissipation.
When comparing Figure \ref{dissipM_dissipZ}a) and b) we see that the effects of accounting for the rotation history of the star are visible, especially during the MS phase.
While the tidal dissipation for a constant spin was constant during most of the MS phase, taking into account the spin down of the star leads to a decrease of the dissipation during this phase. 
In other words, while the dissipation is mainly controlled by the structure during the PMS phase, it is mainly controlled by the rotation on the MS.
The increase of dissipation at the end of the MS phase is still visible but considerably damped due to the much slower rotation.

Figure~\ref{dissipM_dissipZ}b) also shows the effect of the metallicity of the stars on the dissipation within the convective envelope. 
Let us first consider the $1~\Msun$ case.
During the first million years of the PMS phase, the dissipation is higher for lower metallicities and this trend reverses at an age of about 6~Myr.
This is mainly due to the delay of the apparition of the radiative core for higher metallicities.
From that moment on, the dissipation of the metal-rich stars on the MS phase remain much higher (up to 2 orders of magnitude higher) than the dissipation of the metal-poor star.
When the stellar metallicity decreases, and due to opacity effect, the radius and mass of the radiative core tend to increase at a given age. 
As a consequence, and because the size and mass of the convective envelope are reduced, the tidal dissipation strongly decreases. 
This trend is also visible for the $0.4~\Msun$ case, but the difference of the dissipation between the metal-poor star and the two metal-rich stars is much lower.
For all cases, there is a much bigger difference between the Z = 0.004 star and the Z = 0.0134 star than between the Z = 0.0134 star and the Z = 0.0255 star.

The dissipation therefore depends very significantly on the mass of the star and also its metallicity. 
Even for same mass stars, we could therefore expect different planetary tidal orbital evolutions and different stellar rotational evolutions depending on the metallicity.
        

\section{Tidal evolution of planets}

Following the formalism introduced by Bolmont~\&~Mathis~(\cite{Bolmont2016}), we can investigate the effect of stellar mass and metallicity on the tidal evolution of planets.
As Bolmont~\&~Mathis~(\cite{Bolmont2016}) already explored the former for the PMS phase and MS phase, we concentrate here on more advanced stages of evolution: the SG phase and RGB phase.

\subsection{Late evolution of planets}

Bolmont~\&~Mathis~(\cite{Bolmont2016}) showed that the PMS phase has a strong impact on the tidal evolution: Jupiter-mass planets around initially fast rotating stars can migrate outwards to a distance up to 3 times bigger than the initial orbital distance.
We investigate here whether the increase of dissipation (later called the dissipation bump) occurring at the end of the MS phase has an important effect on the tidal evolution of close-in planets.


        \begin{figure}[htbp!]
        \begin{center}
        \includegraphics[width=\linewidth]{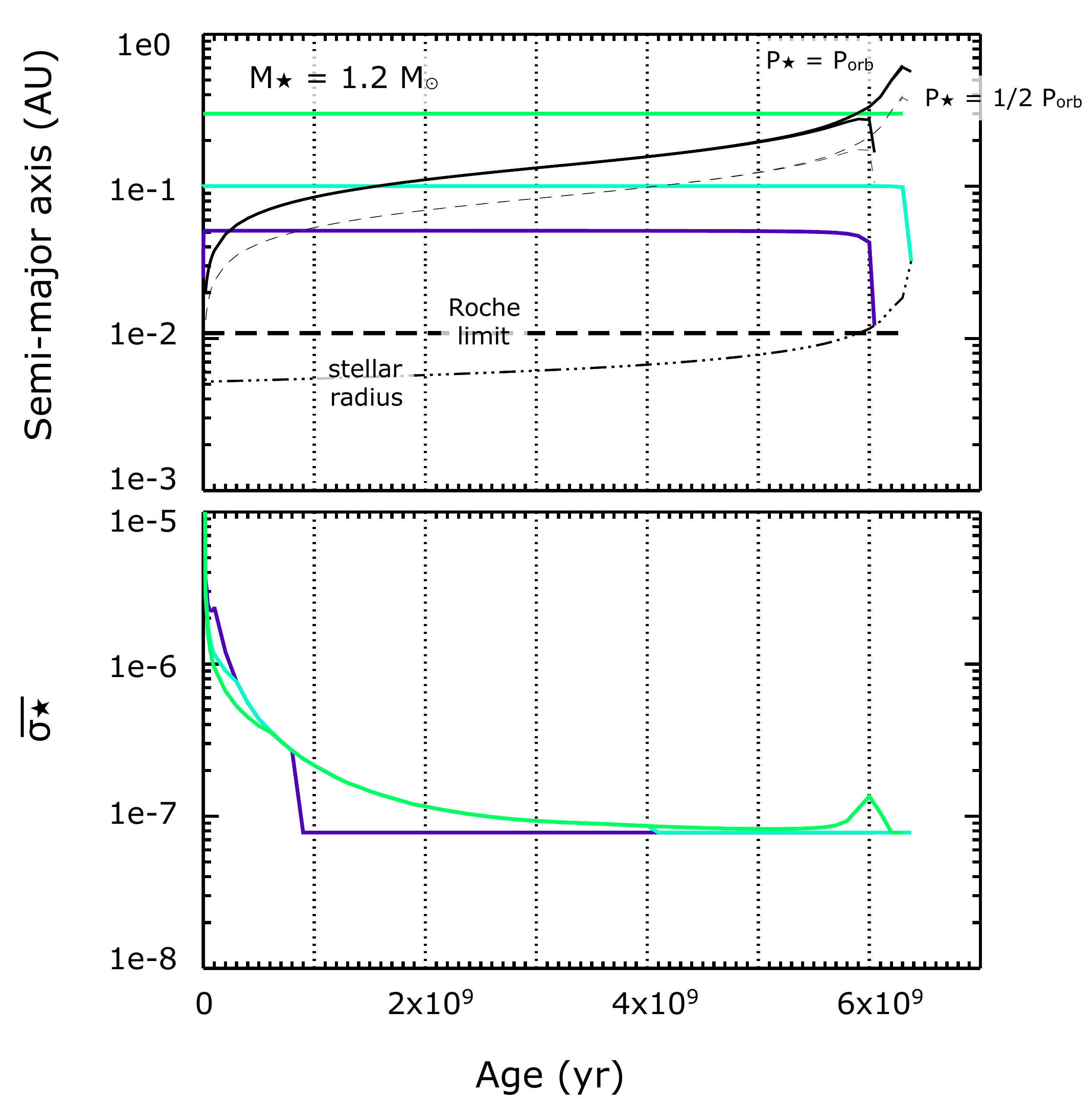}
        \caption{Tidal evolution of a $1~\Mjup$ planet at different initial orbital distances around an initially fast rotating $1.2~\Msun$ star.
        Top panel: Evolution of semi-major axis (in color), $\Ps = \Porb$ (corotation radius, full black line), $\Ps = 1/2~\Porb$ (black light dashed line), the Roche limit (black bold dashed line), the stellar radius (black dashed-dotted line).
        Bottom panel: Evolution of the dissipation factor $\overline{\sigma_\star}$ (defined in Hansen~\cite{Hansen2010}) for the different cases.}
        \label{late_evol_bump}
        \end{center}
        \end{figure}

Figure~\ref{late_evol_bump} shows the evolution of a Jupiter-mass planet around a $1.2~\Msun$ star as well as the evolution of the dissipation factor $\overline{\sigma_\star}$ (defined in Hansen~\cite{Hansen2010}, the higher the dissipation $\left<{\mathcal D}\right>_{\omega}$, the higher $\overline{\sigma_\star}$, see Eqs. 8 and 10 in Bolmont \& Mathis~\cite{Bolmont2016}).
Figure~\ref{late_evol_bump} also shows that the increase of the dissipation occurring at the end of the MS phase has no effect on the tidal evolution of planets.
There are two main reasons for that and both of them are intrinsically linked with the spinning down of the star:
\begin{enumerate}
\item[-] Planets susceptible to feel the effect of the dissipation bump have to be quite far away from the star and thus are less expected to be tidally evolving than closer planets.
Indeed, to feel the effect of the bump, the planet has to be in the region where it excites the inertial tidal waves (i.e., above the black light-dashed line in Figure~\ref{late_evol_bump}, region corresponding to $P_{\rm orb}> 1/2 P_\star$).
As the star spins down due to the stellar winds, this region shrinks in time and the planet has to be far away to be still evolving due to the dynamical tide at the time of the bump ($\sim 6~$Gyr). 
The two closer planets in Figure~\ref{late_evol_bump} are too close to feel the bump: they cross the $\Ps = 1/2~\Porb$ line at $\sim1~$Gyr and $\sim 4$~Gyr and proceed to fall onto the expanding star driven by the equilibrium tide.
However, the farthest planet experiences the late-MS increase of dissipation, but is clearly not impacted.
\item[-] The second reason is that by the time this increase of dissipation occurs, the star has spun down so significantly that the dissipation factor is actually very small.
The bottom panel of Figure~\ref{late_evol_bump} shows that the dissipation corresponding to the bump is only $\sim 2$ times higher than the equilibrium tide dissipation and is actually several orders of magnitude lower than during the PMS phase.   
Due to the effect of the spin on the intensity of the tidal dissipation, the dissipation due to the bump is actually very weak and does not influence the planets susceptible to feel it.
\end{enumerate}

To conclude, as was shown in Bolmont~\&~Mathis~(\cite{Bolmont2016}), the PMS phase leads to the most significant tidal migration.
The late stages of evolution are dominated by the equilibrium tide, which slowly but surely makes the planet spiral inwards on Gyr timescales until they reach the surface of the expanding star.
The planet engulfment leads to an acceleration of the spin of the star as illustrated by the shrinking corotation radius (for the closest planet).
This process may lead to the creation of a population of old fast rotating RGB stars (Privitera~\etal~\cite{Privitera2016}).

\subsection{Influence of the stellar metallicity}

\begin{figure}[htbp!]
        \begin{center}
        \includegraphics[width=\linewidth]{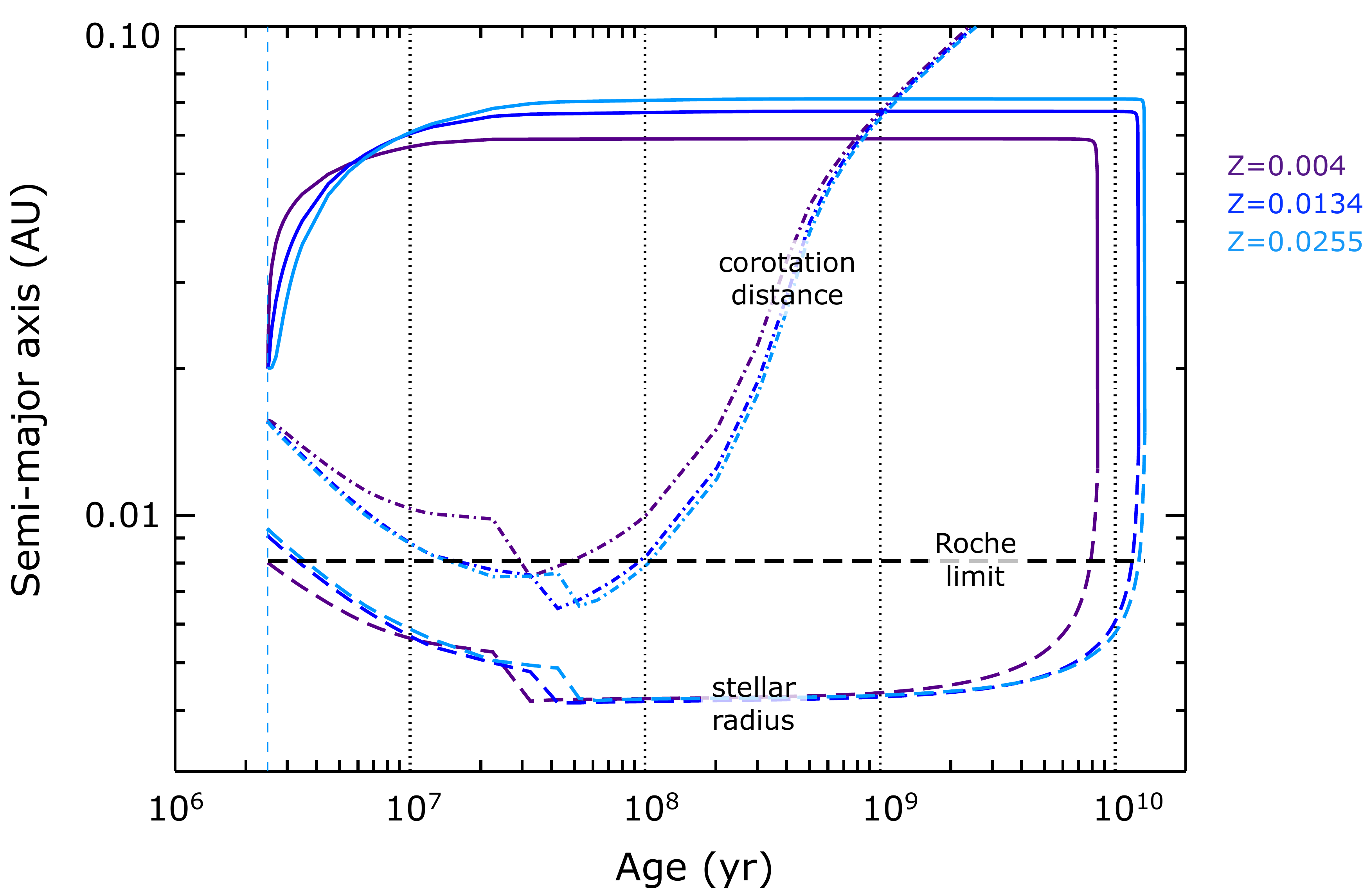}
        \caption{Tidal evolution of a $1~\Mjup$ planet around an initially fast rotating $1~\Msun$ star for different metallicities.
The full colored lines correspond to the semi-major axis evolution of the planets for Z = 0.004 in purple, for Z = 0.0134 in dark blue and for Z = 0.0255 in light blue. 
The corotation distance is represented in colored dashed-dotted lines, the stellar radius in colored long-dashed lines and the Roche limit in black long-dashed lines.}
        \label{sma_diff_Z}
        \end{center}
        \end{figure}

As Figure~\ref{dissipM_dissipZ} shows and as was discussed in Section~\ref{model}, the stellar dissipation in the convective envelope also depends on the stellar metallicity because of the corresponding modification of the stellar structure (Gallet~\etal~in prep).
Figure~\ref{sma_diff_Z} shows the tidal evolution of a Jupiter-mass planet around an initially fast rotating $1~\Msun$ star ($P_{\star, 0} = 1$~day) for different metallicities: Z = 0.004, Z = 0.0134 = $Z_\odot$, and Z = 0.0255. 
The orbital evolution is represented from the PMS phase (from $2.5~$Myr) to the RGB phase. 
As the star is a fast rotator, close-in planets such as represented here are outside the corotation radius. 
The planet therefore migrates outwards in a few $10^7$~yr, here from an initial orbital distance of 0.02~AU to $\sim$0.06 -- 0.07~AU. 

Figure~\ref{sma_diff_Z} also shows that after the initial outward migration the planet orbiting the metal-poor star does not migrate as far as those orbiting more metal-rich stars. 
However in the very beginning of the evolution, it's the opposite: the planet around the metal-poor star migrates outward faster.
This is due to the higher dissipation of the metal-poor star during the first few million years of the PMS phase (see Figure~\ref{dissipM_dissipZ}).
After a few million years, the dissipation of the two metal-rich stars becomes higher and can drive the migration for a longer time.
At the end of the MS, the planet orbiting the metal-poor star is engulfed before the others simply because it is closer and because its radius starts expanding earlier.   

To conclude, planets initially orbiting fast rotating metal-\textit{rich} stars outside the corotation radius migrate farther away than planets around initially fast rotating metal-\textit{poor} stars.

\section{Conclusions}

We investigated the effect of the mass and metallicity of the star on the tidal orbital evolution of close-in Jupiter mass planets.
We found that, while the strong dissipation occurring during the PMS phase has a strong influence on the planets' orbital evolution (see Bolmont~\&~Mathis~\cite{Bolmont2016}), the increase of dissipation at the end of the MS is not responsible for any migration whatsoever.
This is due to the fact that the star considerably spins down from the PMS phase to the RBG phase.

We also found that for initially fast rotating stars, the higher the metallicity of the star is, the farther the planet migrates.
This actually goes against the observations of Adibekyan~\etal~(\cite{Adibekyan2013}) and would necessitate further investigations. 
We should explore the parameter space for the initial conditions of our simulations, and account for the biases that the formation processes induce on the distribution of planets at the end of the protoplanetary disk (i.e., the influence of the metallicity on the initial distribution of planets, see Mordasini~\etal~\cite{Mordasini2012})
Finally, we should investigate the impact of the dissipation occurring in the radiative core, which has been ignored in this study.

\subsection*{Acknowledgments}

E. B. acknowledges that this work is part of the F.R.S.-FNRS ``ExtraOrDynHa'' research project. 
S. M. and E.B. acknowledge funding by the European Research Council through ERC grant SPIRE 647383. 
This work was also supported by the ANR Blanc TOUPIES SIMI5-6 020 01, the Programme National de Plan\'etologie (CNRS/INSU) and PLATO CNES grant at Service d'Astrophysique (CEA-Saclay).
F. G., C. C., and L. A. acknowledge financial support from the SEFRI project C.140049 under COST Action TD 1308 Origins and from the French Programme National National de Physique Stellaire PNPS of CNRS/INSU. 
This work results within the collaboration of the COST Action TD 1308.



\begin{thebibliography}{99}

\bibitem[2013]{Adibekyan2013} Adibekyan, V.~Z., Figueira, P., Santos, N.~C., Mortier, A., Mordasini, C., Delgado Mena, E., Sousa, S.~G., Correia, A.~C.~M., Israelian, G., Oshagh, M.\ 2013.\ Orbital and physical properties of planets and their hosts: new insights on planet formation and evolution.\ Astronomy and Astrophysics 560, A51. 
\bibitem[2016]{Amard2016} Amard, L., Palacios, A., Charbonnel, C., Gallet, F., Bouvier, J.\ 2016.\ Rotating models of young solar-type stars. Exploring braking laws and angular momentum transport processes.\ Astronomy and Astrophysics 587, A105. 
\bibitem[2014]{Auclair-Desrotour2014} Auclair-Desrotour, P., Le Poncin-Lafitte, C., Mathis, S.\ 2014.\ Impact of the frequency dependence of tidal Q on the evolution of planetary systems.\ Astronomy and Astrophysics 561, L7. 
\bibitem[2011]{Bolmont2011} Bolmont, E., Raymond, S.~N., Leconte, J.\ 2011.\ Tidal evolution of planets around brown dwarfs.\ Astronomy and Astrophysics 535, A94. 
\bibitem[2012]{Bolmont2012} Bolmont, E., Raymond, S.~N., Leconte, J., Matt, S.~P.\ 2012.\ Effect of the stellar spin history on the tidal evolution of close-in planets.\ Astronomy and Astrophysics 544, A124. 
\bibitem[2015]{Bolmont2015} Bolmont, E., Raymond, S.~N., Leconte, J., Hersant, F., Correia, A.~C.~M.\ 2015.\ Mercury-T: A new code to study tidally evolving multi-planet systems. Applications to Kepler-62.\ Astronomy and Astrophysics 583, A116. 
\bibitem[2016]{Bolmont2016} Bolmont, E., Mathis, S.\ 2016.\ Effect of the rotation and tidal dissipation history of stars on the evolution of close-in planets.\ Celestial Mechanics and Dynamical Astronomy . 
\bibitem[1997]{Bouvier1997} Bouvier, J., Forestini, M., Allain, S.\ 1997.\ The angular momentum evolution of low-mass stars..\ Astronomy and Astrophysics 326, 1023-1043. 
\bibitem[2008]{Bouvier2008} Bouvier, J.\ 2008.\ Lithium depletion and the rotational history of exoplanet host stars.\ Astronomy and Astrophysics 489, L53-L56. 
\bibitem[2007]{EfroimskyLainey2007} Efroimsky, M., Lainey, V.\ 2007.\ Physics of bodily tides in terrestrial planets and the appropriate scales of dynamical evolution.\ Journal of Geophysical Research (Planets) 112, E12003. 
\bibitem[1998]{EKH1998} Eggleton, P.~P., Kiseleva, L.~G., Hut, P.\ 1998.\ The Equilibrium Tide Model for Tidal Friction.\ The Astrophysical Journal 499, 853-870. 
\bibitem[2015]{GalletBouvier2015} Gallet, F., Bouvier, J.\ 2015.\ Improved angular momentum evolution model for solar-like stars. II. Exploring the mass dependence.\ Astronomy and Astrophysics 577, A98. 
\bibitem[2013]{GalletBouvier2013} Gallet, F., Bouvier, J.\ 2013.\ Improved angular momentum evolution model for solar-like stars.\ Astronomy and Astrophysics 556, A36. 
\bibitem[2009]{GoodmanLackner2009} Goodman, J., Lackner, C.\ 2009.\ Dynamical Tides in Rotating Planets and Stars.\ The Astrophysical Journal 696, 2054-2067. 
\bibitem[2010]{Hansen2010} Hansen, B.~M.~S.\ 2010.\ Calibration of Equilibrium Tide Theory for Extrasolar Planet Systems.\ The Astrophysical Journal 723, 285-299. 
\bibitem[2012]{Hansen2012} Hansen, B.~M.~S.\ 2012.\ Calibration of Equilibrium Tide Theory for Extrasolar Planet Systems. II.\ The Astrophysical Journal 757, 6. 
\bibitem[1981]{Hut1981} Hut, P.\ 1981.\ Tidal evolution in close binary systems.\ Astronomy and Astrophysics 99, 126-140. 
\bibitem[2010]{Leconte2010} Leconte, J., Chabrier, G., Baraffe, I., Levrard, B.\ 2010.\ Is tidal heating sufficient to explain bloated exoplanets? Consistent calculations accounting for finite initial eccentricity.\ Astronomy and Astrophysics 516, A64. 
\bibitem[2013]{Makarov2013} Makarov, V.~V., Efroimsky, M.\ 2013.\ No Pseudosynchronous Rotation for Terrestrial Planets and Moons.\ The Astrophysical Journal 764, 27. 
\bibitem[2009]{Mathis2009} Mathis, S., Le Poncin-Lafitte, C.\ 2009.\ Tidal dynamics of extended bodies in planetary systems and multiple stars.\ Astronomy and Astrophysics 497, 889-910. 
\bibitem[2015]{Mathis2015} Mathis, S.\ 2015.\ Variation of tidal dissipation in the convective envelope of low-mass stars along their evolution.\ Astronomy and Astrophysics 580, L3. 
\bibitem[2016]{Mathis2016} Mathis, S., Auclair-Desrotour, P., Guenel, M., Gallet, F., Le Poncin-Lafitte, C.\ 2016.\ The impact of rotation on turbulent tidal friction in stellar and planetary convective regions.\ Astronomy and Astrophysics 592, A33. 
\bibitem[2015]{Matt2015} Matt, S.~P., Brun, A.~S., Baraffe, I., Bouvier, J., Chabrier, G.\ 2015.\ The Mass-dependence of Angular Momentum Evolution in Sun-like Stars.\ The Astrophysical Journal 799, L23. 
\bibitem[1979]{Mignard1979} Mignard, F.\ 1979.\ The evolution of the lunar orbit revisited. I.\ Moon and Planets 20, 301-315.
\bibitem[2012]{Mordasini2012} Mordasini, C., Alibert, Y., Benz, W., Klahr, H., Henning, T.\ 2012.\ Extrasolar planet population synthesis . IV. Correlations with disk metallicity, mass, and lifetime.\ Astronomy and Astrophysics 541, A97. 
\bibitem[2013]{Ogilvie2013} Ogilvie, G.~I.\ 2013.\ Tides in rotating barotropic fluid bodies: the contribution of inertial waves and the role of internal structure.\ Monthly Notices of the Royal Astronomical Society 429, 613-632. 
\bibitem[2007]{OgilvieLin2007} Ogilvie, G.~I., Lin, D.~N.~C.\ 2007.\ Tidal Dissipation in Rotating Solar-Type Stars.\ The Astrophysical Journal 661, 1180-1191. 
\bibitem[2016]{Privitera2016} Privitera, G., Meynet, G., Eggenberger, P., Vidotto, A.~A., Villaver, E., Bianda, M.\ 2016.\ Star-planet interactions: II. Is planet engulfment the origin of fast rotating red giants?.\ ArXiv e-prints arXiv:1606.08027. 
\bibitem[2012]{Remus2012a} Remus, F., Mathis, S., Zahn, J.-P., Lainey, V.\ 2012.\ Anelastic tidal dissipation in multi-layer planets.\ Astronomy and Astrophysics 541, A165. 
\bibitem[2012]{Remus2012b} Remus, F., Mathis, S., Zahn, J.-P.\ 2012.\ The equilibrium tide in stars and giant planets. I. The coplanar case.\ Astronomy and Astrophysics 544, A132. 
\bibitem[1972]{Skumanich1972} Skumanich, A.\ 1972.\ Time Scales for CA II Emission Decay, Rotational Braking, and Lithium Depletion.\ The Astrophysical Journal 171, 565. 
\bibitem[1989]{Zahn1989} Zahn, J.-P.\ 1989.\ Tidal evolution of close binary stars. I - Revisiting the theory of the equilibrium tide.\ Astronomy and Astrophysics 220, 112-116. 
\bibitem[1977]{Zahn1977} Zahn, J.-P.\ 1977.\ Tidal friction in close binary stars.\ Astronomy and Astrophysics 57, 383-394. 
\bibitem[1975]{Zahn1975} Zahn, J.-P.\ 1975.\ The dynamical tide in close binaries.\ Astronomy and Astrophysics 41, 329-344. 
\bibitem[1966]{Zahn1966} Zahn, J.~P.\ 1966.\ Les mar{\'e}es dans une {\'e}toile double serr{\'e}e (suite).\ Annales d'Astrophysique 29, 489. 





\end{thebibliography}
\end{document}